\documentclass[aps,11pt,nofootinbib,preprintnumbers,groupedaddress,superscriptaddress]{revtex4-2}

\usepackage{amssymb, amsfonts, amsmath, bm}
\usepackage[%dvipdfmx
]{graphicx}
\usepackage{color}
\usepackage{mathrsfs}
\usepackage{enumerate}
\usepackage[table]{xcolor}
\usepackage[%dvipdfmx
]{hyperref}
\hypersetup{%
 colorlinks=true,
  linkcolor=blue,   
  citecolor=blue,   
  urlcolor=blue     
}
\newcommand{\orcid}[1]{%
  \href{https://orcid.org/#1}{\includegraphics[height=0.7em]{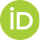}}%
}
\begin{document}
\title{Gravitational lensing and accretion disk imaging of a Buchdahl dense core}
\author{Takahisa Igata\:\!\orcid{0000-0002-3344-9045}
}
\email{takahisa.igata@gakushuin.ac.jp}
\affiliation{
Department of Physics, Gakushuin University,\\
Mejiro, Toshima, Tokyo 171-8588, Japan}
\author{Motoki Omamiuda}
\email{omamiuda1011@gmail.com}
\affiliation{
Department of Physics, Gakushuin University,\\
Mejiro, Toshima, Tokyo 171-8588, Japan}
\author{Yohsuke Takamori\:\!\orcid{0000-0002-2298-195X}}
\email{takamori@wakayama-nct.ac.jp}
\affiliation{National Institute of Technology (KOSEN),\\ 
Wakayama College, Gobo, Wakayama 644-0023, Japan}
\date{\today}

\begin{abstract}
In this paper, we investigate the gravitational lensing and accretion disk imaging characteristics of a dense core modeled by the Buchdahl spacetime. By imposing the appropriate energy conditions and ensuring the absence of curvature singularities, we delineate the parameter space in which the dense core mimics key gravitational features of black holes while exhibiting unique deviations. We derive the photon orbital equation and calculate deflection angles, clearly distinguishing between weak- and strong-deflection regimes. Furthermore, we construct a mapping from the illuminated, geometrically thin accretion disk onto the observer's screen---focusing on the isoradial curves corresponding to a representative source ring. For compactness values below a critical threshold, only a finite number of disk images are formed. In this range, their secondary and higher-order images typically display double-loop structures, with each loop individually capturing the entire source ring. Notably, the highest-order image sometimes appears as a single, crescent-shaped loop that does not enclose the screen's center, implying the existence of a cutoff angle that restricts the imaged portion of the source ring. In contrast, for compactness values above the critical threshold, an infinite sequence of double-loop structures appears---a behavior closely linked to the presence of a photon sphere. These findings suggest that the lensing signatures of dense cores can distinguish them from black holes, offering new insights for high-resolution observations.
\end{abstract}
\maketitle

%%%%%%%%%%
\section{Introduction}
\label{sec:1}
%%%%%%%%%%
Over the past decade, remarkable advances in observational astronomy have significantly deepened our understanding of the compact objects at the centers of galaxies, providing new insights into their gravitational effects. Among these objects are M87* and Sagittarius A*, which are widely regarded as supermassive black hole candidates and have attracted considerable attention from theoretical and observational perspectives. The groundbreaking imaging of these objects by the Event Horizon Telescope~\cite{EventHorizonTelescope:2019dse,EventHorizonTelescope:2022wkp} provided compelling evidence in support of the existence of a black hole, or at least an object with black-hole-like properties. The sharp boundaries of the observed shadows likely arise from both the strong bending of light near the photon sphere and the potential absorption near the Schwarzschild radius, each of which is indicative of an intense gravitational field consistent with black hole predictions. Nevertheless, confirming the existence of a black hole and its event horizon is essentially impossible, as the event horizon represents a boundary from which neither light nor matter can escape~\cite{Hawking:1973uf}, rendering direct observational verification unattainable~\cite{Abramowicz:2002vt,Nakao:2018knn}. Thus, while the observed shadows align well with black hole models, they neither conclusively prove the presence of a horizon nor rule out the possibility of alternative objects.

This fundamental limitation---the inability to definitively confirm the existence of an event horizon---has motivated extensive research into horizonless compact objects~\cite{Cardoso:2017cqb}, including naked singularities, wormholes, boson stars, gravastars, and dense cores. These objects exhibit strong gravitational fields comparable to those of black holes, but could, in principle, allow the direct observation of their interiors or surfaces. Observationally, one promising approach is to look for small yet significant deviations by comparing the photon ring and shadow features predicted by standard black hole models with those observed in data. If such discrepancies are found, they would substantially strengthen the possibility that the central object is an alternative to a black hole, thereby broadening our understanding of supermassive objects at galactic centers. Theoretically, it is critical to develop and refine alternative models that quantitatively predict such observational signatures. Systematic comparisons between these predictions and high-precision data could determine whether alternative objects mimic all black hole observables or reveal detectable differences. Even if no observational test can definitively rule out the standard black hole model, continued efforts in model building and data analysis promise invaluable insights into the nature of compact objects at galactic centers. Consequently, determining whether a genuinely alternative configuration can replicate every black hole signature---or reveal measurable deviations---remains an open and compelling question. In this context, upcoming developments in very-long-baseline interferometry (e.g., Ref.~\cite{Lupsasca:2024xhq}) will further refine angular resolution, offering new opportunities to test such scenarios and explore possible deviations from the Kerr metric~\cite{Cardoso:2019rvt}.

Given these considerations, imaging of accretion disks plays a central role in understanding the gravitational environments surrounding compact objects and serves as a powerful tool for distinguishing between competing theoretical models. Early studies~\cite{Cunningham:1973, Luminet:1979, Fukue:1988, Takahashi:2004xh} demonstrated how the gravitational field of a black hole strongly bends light emitted by an accretion flow, producing observable features such as distorted disk images and photon rings. More recent studies have explored thin accretion disk images around black holes in various contexts~\cite{Tsupko:2022kwi, Liu:2021lvk, Guo:2023zwy, Gyulchev:2021dvt,Zhu:2024vxw}. Meanwhile, researchers have investigated the optical appearance of a thin accretion disk surrounding a black hole alternative, including naked singularities~\cite{Gyulchev:2019tvk, Gyulchev:2020cvo, Guerrero:2022msp}, wormholes~\cite{Paul:2019trt, Rahaman:2021kge}, boson stars~\cite{Rosa:2023qcv}, and gravastars~\cite{Rosa:2024bqv}, revealing how the underlying geometry can imprint distinctive observational signatures. Collectively, these studies underscore the power of disk imaging to uncover subtle yet critical differences between standard black holes and potential alternatives. 

Among the many alternatives proposed, dense cores have garnered particular attention because they can remain extremely compact---reproducing the key gravitational effects of black holes---while maintaining a smooth and physically consistent matter distribution. Recent studies on the core-halo distribution of dark matter suggest that dense cores may account for the environments in galactic centers~\cite{Ruffini:2014zfa, Arguelles:2016ihf}, and relativistic effects such as periapsis shifts in dense core models have been analyzed~\cite{Rubilar:2001, Igata:2022nkt}. Moreover, distinct features in photon ring structures or luminosity profiles arising from a dense core's internal mass distribution could distinguish it from a black hole. For instance, some numerical analyses of dark matter cores illuminated by accretion disks report central brightness depressions without clear photon rings, highlighting key distinctions from black hole scenarios~\cite{Pelle:2024eyt}. 

Despite this growing interest, systematic studies of optically thin accretion disks around dense cores, employing analytically tractable models, remain limited. Dense core solutions vary widely in their matter distributions and underlying assumptions, and purely numerical models may obscure important physical insights. Building on prior developments, this paper aims to provide new perspectives on the gravitational lensing and observational signatures of dense cores by adopting a simpler perfect-fluid solution. This approach clarifies how the compactness and mass distribution of a dense core shape its associated lensing effects and disk images, thereby highlighting potential observational signals that could differentiate dense cores from black holes. Such signals may be investigated in future high-resolution observations, offering deeper insights into the nature of these compact objects.

In particular, this study focuses on dense cores modeled by the Buchdahl solution, a static and spherically symmetric perfect-fluid configuration~\cite{Buchdahl:1964}. The Buchdahl solution serves as a relativistic extension of the Newtonian Plummer model, providing a useful bridge between Newtonian gravity and strong gravity regimes. Unlike naked singularities, this solution describes a regular spacetime that is both highly compact and free of singularities. Furthermore, whereas many alternative black hole models invoke exotic matter fields, the Buchdahl solution relies on a classical perfect fluid that satisfies standard energy conditions. This combination of regularity and physical plausibility makes Buchdahl-like dense cores an ideal model for studying compact objects in a near-black-hole regime, demonstrating how they may reproduce or deviate from typical black hole signatures. 

This paper is organized as follows: In Sec.~\ref{sec:2}, we review the Buchdahl spacetime and summarize the conditions under which it remains physically viable. In Sec.~\ref{sec:3}, we establish the photon trajectory equation and derive the deflection angle, providing a quantitative measure of gravitational lensing. In Sec.~\ref{sec:4}, we present the formulation of disk imaging in the Buchdahl core, introducing the concept of isoradial curves. In Sec.~\ref{sec:5}, we report our disk imaging results, illustrating how the disk's appearance depends on the core's compactness and the observer's inclination. Finally, in Sec.~\ref{sec:6}, we discuss the broader implications of our findings in the context of gravitational lensing and astrophysical observations.

Throughout this work, we employ geometrized units, setting the gravitational constant $G$ and the speed of light $c$ to unity (i.e., $G=1$ and $c=1$).

%%%%%%%%%%
\section{Buchdahl solution: metric, energy conditions, and physical interpretation}
\label{sec:2}
%%%%%%%%%%
The Buchdahl spacetime~\cite{Buchdahl:1964} is a static, spherically symmetric solution to Einstein's field equations, often considered as a theoretical model for dense cores. In isotropic coordinates, the metric $g_{ab}$ is given by the line element
\begin{align}
\mathrm{d}s^2
&=-\frac{(1-f)^2}{(1+f)^2}\:\!\mathrm{d}t^2+(1+f)^4\left(
\mathrm{d}r^2+r^2\:\!\mathrm{d}\theta^2+r^2\sin^2\theta\:\!\mathrm{d}\varphi^2
\right),
\end{align}
with 
\begin{align}
f(r)=\frac{a}{2\sqrt{1+k r^2}},
\label{eq:f}
\end{align}
where $t$ is the time coordinate for static observers, $(r, \theta, \varphi)$ are the standard spherical coordinates, and $a$ and $k$ are constants (their physical interpretations are discussed later). To ensure a physically meaningful spacetime, we assume $a\neq 0$ and $k\neq 0$. The case $a=0$ corresponds to a flat metric, while $k=0$ results in either a degenerate or singular metric when $|a|= 2$, or a flat metric otherwise. 

The Kretschmann scalar $K\equiv R_{abcd}R^{abcd}$ for the Buchdahl spacetime can be written as 
\begin{align}
K=\frac{k^2 P_n(f)}{(1-f)^2(1+f)^{n}},
\end{align}
where $n=12$ for $|a|\neq 2$ and $n=10$ for $|a|=2$. Here, the function $P_n(f)$ is a polynomial in $f$ that remains finite at $|f|=1$. The Kretschmann scalar $K$ diverges as $K \propto (1-f)^{-2}$ when $f\to 1$ and as $K \propto (1+f)^{-n}$ when $f\to -1$. These divergences indicate the presence of curvature singularities at the points where $|f|=1$, which must be excluded to ensure that spacetime remains physically viable.

To maintain the regularity of the spacetime, the function $f$ imposes constraints on the parameters $a$ and $k$. The range of $f$ and the associated constraints, determined by the signs of $a$ and $k$, can be summarized as follows:

When $a>0$ and $k>0$, the range of $f$ is $(0, a/2\:\!]$. 
Ensuring that $f\neq1$ requires $0<a<2$. 

When $a>0$ and $k<0$, the range of $f$ is $[\:\!a/2, \infty)$. 
Ensuring that $f\neq1$ requires $a>2$. 

When $a<0$ and $k>0$, the range of $f$ is $[\:\!a/2, 0)$. 
Ensuring that $f\neq-1$ requires $-2<a<0$. 

When $a<0$ and $k<0$, the range of $f$ is $(-\infty, a/2\:\!]$. 
Ensuring that $f\neq-1$ requires $a<-2$. 

\noindent
From these considerations, the Buchdahl solution remains free of curvature singularities in the parameter ranges $0<|a|<2$ with $k>0$, or $|a|>2$ with $k<0$.

By applying Einstein's field equations, one obtains the stress-energy tensor in the perfect-fluid form,
\begin{align}
T_{ab}&=\rho(r) u_a u_b+p(r) (g_{ab}+u_a u_b),
\label{eq:stress-energy tensor}
\end{align}
with
\begin{align}
\rho(r)&=\frac{24 k f^5}{\pi a^4 (1+f)^5},
\label{eq:rho}
\\
p(r)&= \frac{8k f^6}{\pi a^4(1-f)(1+f)^5}, 
\label{eq:p}
\end{align}
where $\rho(r)$ is the energy density, $p(r)$ is the pressure, and 
$u^a$ is the four-velocity of static observers satisfying $u^au_a=-1$. 

Given that the stress-energy tensor \eqref{eq:stress-energy tensor} is that of a perfect fluid, the standard energy conditions can be stated as follows: 
\begin{enumerate}[(i)\,]
\item The weak energy condition (WEC) requires that $\rho \ge0$ and $\rho+p\ge0$.
\item The strong energy condition (SEC) requires that $\rho+3p \ge 0$ and $\rho+p\ge0$.
\item The null energy condition (NEC) requires that $\rho+p\ge 0$.
\item The dominant energy condition (DEC) requires that $\rho\ge |p|$. 
\end{enumerate}

We now seek values of $a$ and $k$ for which all energy conditions (i)--(iv) hold in a singularity-free spacetime. In particular, we focus on the range $0<a<2$ and $k>0$ (see the Appendix for other ranges). In this regime, the function $f$ satisfies $f\in(0,a/2\:\!]$,
and the energy density \eqref{eq:rho} remains strictly positive ($\rho>0$) everywhere.

Since $\rho>0$, the WEC and NEC require $p/\rho\ge -1$, 
while the SEC imposes $p/\rho\ge -1/3$. 
From Eqs.~\eqref{eq:rho} and \eqref{eq:p}, 
the ratio $p/\rho$ is given by
\begin{align}
\frac{p}{\rho}=\frac{f}{3(1-f)}.
\end{align}
For $f\in(0,a/2\:\!]$, 
this relation implies that $0<p/\rho <a/[3(2-a)]$.
Since these values meet the WEC, NEC, and SEC requirements, all three
conditions are satisfied throughout the spacetime. 

The DEC requires $|p|/\rho \le 1$, equivalently,
\begin{align}
\bigg|\frac{p}{\rho}\bigg|=\bigg|\frac{f}{3(1-f)}\bigg|\le 1.
\end{align}
This condition is satisfied if $f\le 3/4$. Since $f\le a/2$ within the specified parameter range, the DEC holds throughout the spacetime, provided that $a\leq 3/2$.

As summarized in Table~\ref{table:EC}, there are no other parameter ranges that satisfy all of the energy conditions across the entire spacetime. We therefore focus our analysis on 
\begin{align}
0<a\leq\frac{3}{2}, \quad k>0.
\end{align}

\begin{table}[t]
\begin{tabular}{cccccccc}
\hline
\hline
&&~~~$a<-2$~~~&~~$-2<a<0$~~&~~$0<a\leq3/2$~~&~~$3/2<a<2$~~&~$2<a\le3$~&~~$a>3$~~
\\
\hline
~$k>0$~~&
\begin{tabular}{l}
~~WEC~~\\~~NEC~~\\~~SEC~~\\~~DEC~~
\end{tabular}
&\cellcolor{gray!10}(Singular)&
\begin{tabular}{c}
$\times$\\$\times$\\$\times$\\$\times$
\end{tabular}
&
\begin{tabular}{c}
$\checkmark$\\$\checkmark$\\$\checkmark$\\$\checkmark$
\end{tabular}
&
\begin{tabular}{c}
$\checkmark$\\$\checkmark$\\$\checkmark$\\$\times$
\end{tabular}
&
\cellcolor{gray!10}(Singular)
&
\cellcolor{gray!10}(Singular)
\\
\hline
~$k<0$~~&
\begin{tabular}{l}
~~WEC~~\\~~NEC~~\\~~SEC~~\\~~DEC~~
\end{tabular}
&
\begin{tabular}{c}
$\times$\\$\times$\\$\times$\\$\times$
\end{tabular}
&\cellcolor{gray!10}(Singular)
&\cellcolor{gray!10}(Singular)
&\cellcolor{gray!10}(Singular)&
\begin{tabular}{c}
$\times$\\$\mathrm{P}$\\$\checkmark$\\$\times$
\end{tabular}
&
\begin{tabular}{c}
$\times$\\$\times$\\$\checkmark$\\$\times$
\end{tabular}
\\
\hline\hline
\end{tabular}
\caption{Summary of spacetime regularity and energy conditions (WEC, NEC, SEC, and DEC) across parameter ranges for $a$ and $k$.
``Singular" denotes the case where curvature singularities appear. Here, $\checkmark$ indicates that the energy condition is satisfied, $\times$ indicates that it is violated, and $\mathrm{P}$ indicates that the condition is partially satisfied.}
\label{table:EC}
\end{table}

In the Buchdahl spacetime, the areal radius $\tilde{r}$ is related to the isotropic radial coordinate $r$ by
\begin{align}
\tilde{r}(r)=r(1+f)^2.
\end{align}
In the asymptotic region ($\sqrt{k}r \gg 1$), the isotropic radial coordinate $r$ becomes nearly equal to the areal radius $\tilde{r}$ (i.e., $r \simeq \tilde{r}$). 
The Misner--Sharp mass $m(\tilde{r})$ is defined as
\begin{align}
m(r)=\frac{\tilde{r}}{2} [1-g^{ab}(\nabla_a \tilde{r})(\nabla_b \tilde{r})], 
\end{align}
where $g^{ab}$ denotes the inverse of $g_{ab}$. This quantity serves as a quasilocal measure of the gravitational mass enclosed within the areal radius $\tilde{r}$~\cite{Misner:1964je, Hayward:1994bu, Kinoshita:2024wyr}. As a result of an explicit calculation, the Misner--Sharp mass is given by 
\begin{align}
m(r)=-2r^2 f'(1+f+ rf'),
\label{eq:MS}
\end{align}
where the prime denotes differentiation with respect to $r$. 
The total mass $M$ of the spacetime is given by
\begin{align}
M=\lim_{r\to \infty} m(r)=
\frac{a}{\sqrt{k}},
\label{eq:M}
\end{align}
which corresponds to the ADM mass.

We now clarify the physical interpretation of the parameters $a$ and $k$. 
The parameter $k$ sets a characteristic length scale in the spacetime, and we define the core radius $R$ by 
\begin{align}
R=\frac{1}{\sqrt{k}}.
\label{eq:R}
\end{align}
Substituting $r=R$ into Eq.~\eqref{eq:MS} and using Eq.~\eqref{eq:M}, we obtain
\begin{align}
m(R)=\frac{M}{\sqrt{2}}\approx 0.707 M. 
\end{align}
This result suggests that approximately $70.7\%$ of the total mass $M$ is enclosed within the radius $r=R$, thereby supporting the interpretation of $R$ as the core radius of the matter distribution. 
From Eqs.~\eqref{eq:M} and \eqref{eq:R}, we see that $a$ is the dimensionless ratio of $M$ to $R$, 
\begin{align}
a=\frac{M}{R}. 
\end{align}
Larger values of $a$ indicate a more compact object with a stronger gravitational field. For this reason, we refer to $a$ as the compactness parameter of the core.

%%%%%
\section{Gravitational lensing in the Buchdahl spacetime}
\label{sec:3}
%%%%%
The dynamics of photons in the Buchdahl spacetime is described by the following Lagrangian:
\begin{align}
\mathscr{L}
=\frac{1}{2}\left[\:\!
-\frac{(1-f)^2}{(1+f)^2} \dot{t}^2+(1+f)^4\left(\dot{r}^2+r^2\dot{\theta}^2+r^2 \sin^2\theta\:\!\dot{\varphi}^2\right)
\:\!\right],
\end{align}
where the dot denotes differentiation with respect to an arbitrary dimensionless affine parameter $\lambda$. 
Due to the spherical symmetry of the spacetime, photon trajectories can be restricted to the equatorial plane (i.e., $\theta=\pi/2$) without loss of generality. Under this assumption, since $\theta=\pi/2$ (implying that $\dot{\theta}=0$), the Lagrangian reduces to 
\begin{align}
\mathscr{L}
=\frac{1}{2}\left[\:\!
-\frac{(1-f)^2}{(1+f)^2} \dot{t}^2+(1+f)^4\left(\dot{r}^2+r^2 \:\!\dot{\varphi}^2\right)
\:\!\right].
\end{align}

Owing to the static and axial symmetries of the spacetime, two conserved quantities arise. Time-translation invariance implies the conservation of photon energy, 
\begin{align}
E=-\frac{\partial \mathscr{L}}{\partial \dot{t}}
=\frac{(1-f)^2}{(1+f)^2}\dot{t},
\end{align}
while rotational invariance implies the conservation of photon angular momentum,
\begin{align}
L=\frac{\partial \mathscr{L}}{\partial \dot{\varphi}}
=r^2(1+f)^4 \dot{\varphi}. 
\end{align}

Substituting these constants of motion back into the Lagrangian and imposing the null condition $\mathscr{L}=0$ yields the radial equation of motion,
\begin{align}
\frac{(1-f)^2}{r^4(1+f)^6} \left(\frac{\mathrm{d}r}{\mathrm{d}\varphi}\right)^2+V(r)=\frac{1}{b^2},
\end{align}
where $b=L/E$ denotes the impact parameter, and the effective potential is defined as 
\begin{align}
V(r)=\frac{(1-f)^2}{r^2(1+f)^6}.
\label{eq:V}
\end{align}

The effective potential $V(r)$ strongly depends on the compactness parameter $a$. 
Specifically, the analysis of photon circular orbits in Ref.~\cite{Igata:2022nkt} yields the critical value,
\begin{align}
a_0=1.3629,
\end{align}
which is obtained by simultaneously imposing the conditions $V'(r)=0$ and $V''(r)=0$. This ensures that the potential has an inflection point where its curvature changes sign. If $a\ge a_0$, the potential $V(r)$ exhibits both a local maximum and a local minimum, 
corresponding to unstable and stable photon circular orbits, respectively. At the radii corresponding to these extrema, one obtains the photon sphere and the antiphoton sphere~\cite{Gibbons:2016isj,Cunha:2017qtt,Kudo:2022ewn}, respectively. For $a<a_0$, the potential decreases monotonically, implying the absence of such orbits.

\begin{figure}[t]
\centering
\includegraphics[width=8cm,clip]{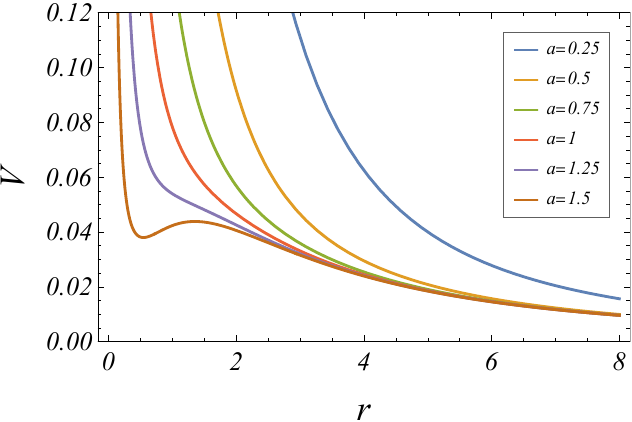}
\caption{The effective potential $V(r)$ as a function of the radial coordinate $r$ is shown for different values of the compactness parameter $a$, with units set to $M=1$. 
}
\label{fig:effective_potential}
\end{figure}

Figure~\ref{fig:effective_potential} illustrates $V(r)$ as a function of the radial coordinate $r$ for various values of $a$. For instance, when $a=1.5$ ($a > a_0$), the effective potential clearly exhibits both a local maximum and a local minimum. In contrast, for $a < a_0$, $V(r)$ decreases monotonically.
%%%%%
%%%%%

Next, we introduce the inverse radial coordinate $u=1/r$ (so that $f$ is now understood as $f(1/u)$), which facilitates rewriting the radial equation as
\begin{align}
\left(\frac{\mathrm{d}u}{\mathrm{d}\varphi}\right)^2=
G(u; b)
\label{eq:EOM}
\end{align}
where 
\begin{align}
G(u; b)=\frac{1}{b^2}\frac{(1+f)^6}{(1-f)^2}-u^2.
\end{align}
To relate the impact parameter $b$ to the closest approach distance $P$, 
we examine the point where the radial derivative vanishes, i.e., $\mathrm{d}u/\mathrm{d}\varphi|_{u=1/P}=0$. Substituting $u=1/P$ into Eq.~\eqref{eq:EOM}, we obtain $G(P^{-1}; b)=0$. This condition yields the relation between $b$ and $P$ as
\begin{align}
b^2=P^2 \frac{(1+f(P))^6}{(1-f(P))^2}. 
\end{align}

\begin{figure}[t]
\centering
\includegraphics[width=8cm,clip]{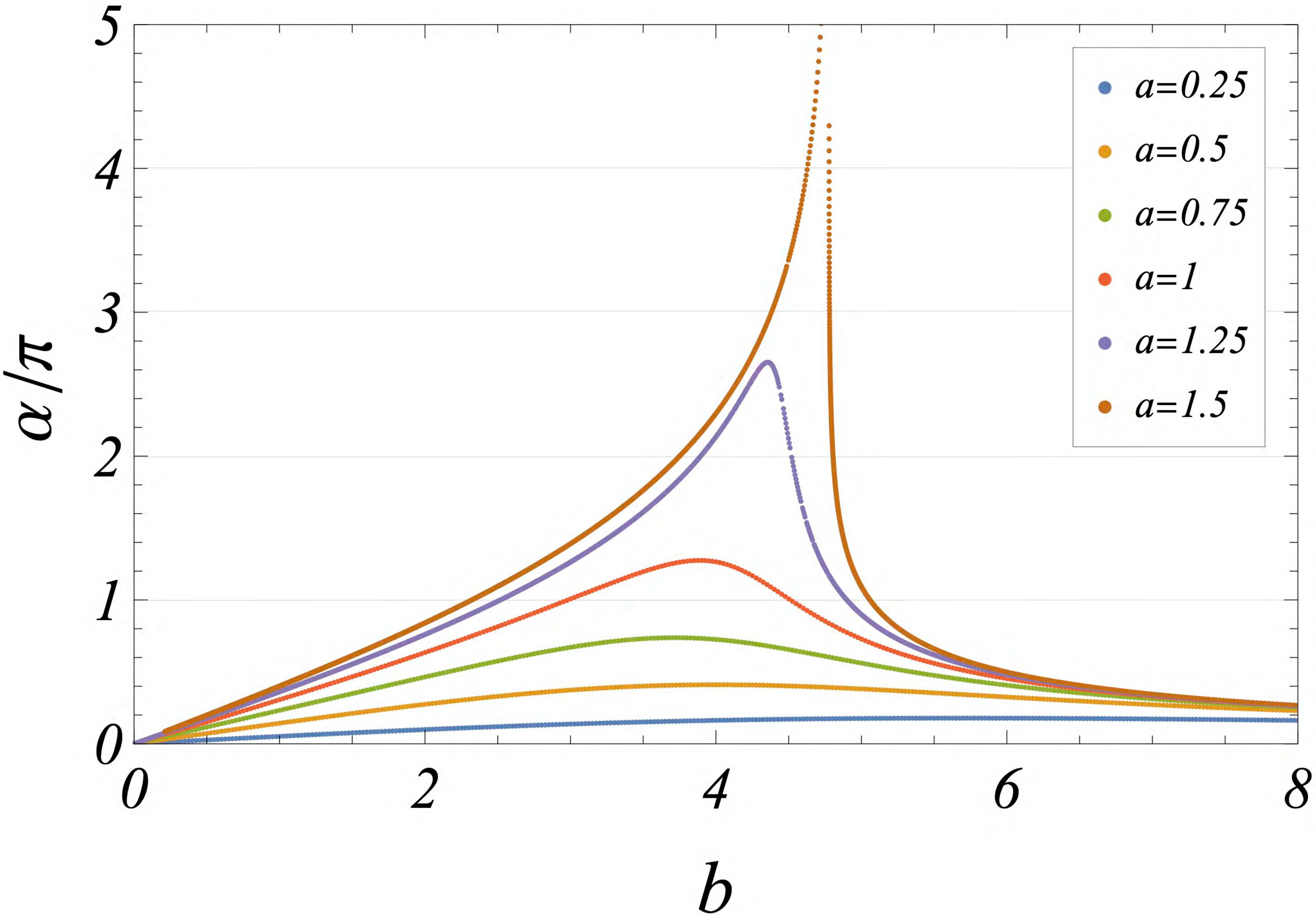}
\caption{The deflection angle $\alpha$ (normalized by $\pi$) as a function of $b$ for various values of the compactness parameter $a$, with units set to $M=1$.
Horizontal lines indicate integer multiples of $\pi$, representing the number of half-windings completed by the photons. 
}
\label{fig:deflection}
\end{figure}

The deflection angle $\alpha$ for a photon with impact parameter $b$ is given by
\begin{align}
\alpha=2 \int_0^{1/P} \frac{\mathrm{d}u}{\sqrt{G(u;b)}} -\pi. 
\label{eq:deflectionangle}
\end{align}
Figure~\ref{fig:deflection} presents the normalized deflection angle $\alpha/\pi$ as a function of $b$ for various values of the compactness parameter $a$. Each curve exhibits either a single peak or a divergence. In both cases, $\alpha/\pi$ decreases monotonically as $b$ moves away from a characteristic value. Determining whether the peak exceeds $\alpha/\pi=1$ or diverges entirely is crucial for distinguishing between different deflection regimes and for understanding strong gravitational lensing effects in this spacetime.
%%%%%
%%%%%

We classify photon deflection behavior into two regimes: weak and strong. 
The boundary between these regimes occurs at $\alpha=\pi$. 
In the weak-deflection regime ($\alpha<\pi$), photons experience relatively small angular deviations. 
In contrast, in the strong-deflection regime ($\alpha> \pi$), photons are deflected by more than $\pi$ before escaping, indicating 
a significantly stronger gravitational influence on their trajectories. 
The onset of strong deflection depends on both the compactness parameter $a$ and the impact parameter $b$. 
Additionally, within the strong-deflection regime, we distinguish between two cases based on whether an unstable photon circular orbit is present. 

For small values of the compactness parameter ($a\ll1$), the deflection angle remains below $\pi$ for all $b$, 
implying that the gravitational field of the dense core is too weak to generate strong deflection. 

For intermediate values of the compactness parameter ($1 \lesssim a<a_0$), a finite range of $b$ values results in strong deflection, allowing photons to orbit the dense core multiple times before escaping. Outside this range, weak deflection dominates.
As $a$ increases in this intermediate regime, the maximum possible number of photon windings increases, highlighting how the compactness parameter enhances gravitational lensing effects. 

When the compactness reaches or exceeds the critical value $a\ge a_0=1.3629$, strong deflection occurs within a finite range of $b$. However, $\alpha/\pi$ diverges as $b$ approaches a specific critical value. This divergence is associated with the presence of an unstable photon circular orbit, which allows photons to orbit the dense core an arbitrarily large number of times before escaping. 
Such behavior resembles the characteristic signature of a photon sphere, which is well known in black hole spacetimes~\cite{Bozza:2002zj,Tsukamoto:2016jzh,Shaikh:2019itn,Tsukamoto:2020iez}. 

Beyond the general characterization of the strong deflection regime,
Fig.~\ref{fig:deflection} also illustrates the formation of multiple images, produced by photons that orbit the dense core multiple times. 
Let $N$ be an integer with $N\ge 1$.
If the interval $N\le \alpha/\pi<N+1$ corresponds to a single continuous range of $b$, 
the resulting image will likely appear as a single connected structure. 
Conversely, if $N\le \alpha/\pi<N+1$ is realized for two disjoint intervals of $b$, 
then the corresponding image may split into two disconnected components. 
This subtle relationship between $b$ and image topology can have significant consequences for gravitational lensing observations.

In summary, strong deflection in the Buchdahl spacetime occurs under specific conditions determined by both the compactness parameter $a$ and the impact parameter $b$.
The compactness parameter $a$ plays a crucial role in governing the strength and qualitative nature of the bending, while the critical value $a_0$ marks the transition between finite and diverging deflection angles. These results underscore how $a$ regulates the lensing phenomenology, affecting not only the degree of photon deflection but also the complexity of multiple images.

%%%%%
\section{Formulation of disk imaging in the Buchdahl dense core}
\label{sec:4}
%%%%%
We consider a geometrically thin accretion disk that encircles a dense core (modeled by the Buchdahl solution) and emits radiation detectable at large distances. Specifically, we assume that the disk's vertical thickness is negligible compared to the characteristic length scale set by the mass $M$. This idealization simplifies the overall geometry, thereby facilitating the interpretation of the disk's apparent image. Note that photons emitted by the disk propagate through the curved spacetime around or within the dense core, and only a fraction of these photons ultimately reach the observer. 

%%%%%
\begin{figure}[t]
\centering
\includegraphics[width=10cm,clip]{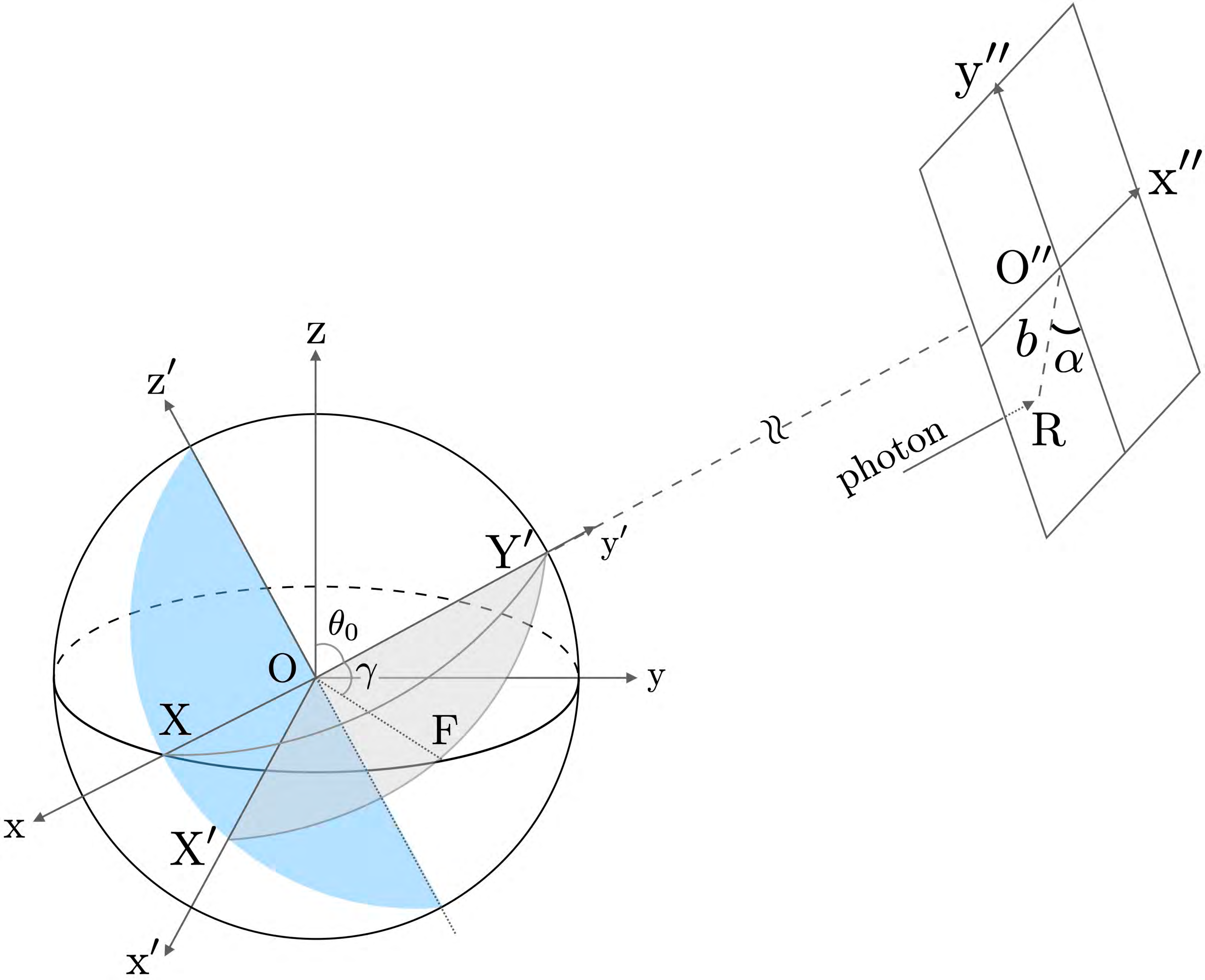}
\\
\caption{Schematic view of the geometric configuration. The dense core is centered at $\mathrm{O}$, and the accretion disk lies on the $z=0$ plane. A photon emitted from an emission point eventually crosses the equatorial plane at the final crossing point $\mathrm{F}$ and reaches $\mathrm{R}$ on the observer's screen. The gray-shaded region represents the photon's orbital plane. The blue-shaded region is parallel to the observer's screen.}
\label{fig:configuration}
\end{figure}
%%%%%

We largely follow the formalism presented in Ref.~\cite{Luminet:1979}, with slight modifications to certain definitions to simplify our analysis. As illustrated in Fig.~\ref{fig:configuration}, the observer is located at $\mathrm{O}''$ with
spatial coordinates $(r,\theta,\varphi)=(r_0, \theta_0, 0)$, where $r_0\gg M$. The observer's screen---oriented perpendicular to the line of sight connecting $\mathrm{O}''$ and the center $\mathrm{O}$ of the dense core---serves as the reference plane for visualizing the accretion disk.

The accretion disk lies on the equatorial plane $\theta=\pi/2$. Consider a photon emitted from a point $\mathrm{S}$ on the disk with spherical coordinates $(r, \pi/2, \varphi)$. This photon follows a spatial trajectory $\mathcal{C}$ and eventually reaches the observer's screen at $\mathrm{R}$. To specify the location of $\mathrm{R}$ on the screen, we introduce Cartesian coordinates $(x'', y'')$ centered at $\mathrm{O}''$. The $\mathrm{y}''$ axis is oriented in the direction of decreasing $\theta$ at $\mathrm{O}''$, while the $\mathrm{x}''$ axis is oriented in the direction of increasing $\varphi$ (note that these directions differ from those used in Ref.~\cite{Luminet:1979}). Alternatively, $\mathrm{R}$ can be described in polar coordinates $(b, \alpha)$, where $b$ is the photon's impact parameter, and $\alpha$ is measured clockwise from the negative $\mathrm{y}''$ axis. The coordinate transformation is then given by%
\footnote{Note that $\alpha$ introduced here should not be confused with the deflection angle defined in Eq.~\eqref{eq:deflectionangle}.}
\begin{align}
(x'', y'')=(-b \sin \alpha, -b \cos \alpha),
\end{align}
with $\alpha$ restricted to $-\pi \le \alpha \le \pi$.

We now derive the total change in the orbital angle---i.e., the angle measured in the photon's orbital plane---along its trajectory $\mathcal{C}$, using a dynamical approach. Note that this change, denoted by $\Delta \tilde{\varphi}$, should not be confused with the change in the global angular coordinate $\varphi$. In this framework, if no turning point occurs along $\mathcal{C}$, we have
\begin{align}
\Delta \tilde{\varphi}=\int_0^{1/r} \frac{\mathrm{d}u}{\sqrt{G(u;b)}}.
\label{eq:Deltavarphi1}
\end{align}
If a turning point $\mathrm{p}$ exists (with radial coordinate $P$ corresponding to the point of closest approach to the core center), then
\begin{align}
\Delta \tilde{\varphi}=\int_{0}^{1/r} \dfrac{\mathrm{d}u}{\sqrt{G(u;b)}}+2 \int_{1/r}^{1/P} \frac{\mathrm{d}u}{\sqrt{G(u;b)}}.
\label{eq:Deltavarphi2}
\end{align}
Since the photon is emitted from $\mathrm{S}$ on the equatorial plane, its azimuthal change can be naturally decomposed as 
\begin{align}
\Delta \tilde{\varphi}=n\pi+\gamma,
\label{eq:deltavarphigamma}
\end{align}
where $n$ is a nonnegative integer representing the number of times the photon crosses the equatorial plane (with $n=0$ at $\mathrm{S}$), and $\gamma$ represents the residual angular deviation beyond these complete half-turns (i.e., the additional bending angle, with $0\le \gamma<\pi$). In what follows, we relate $\gamma$ to the observed screen angle $\alpha$, thereby establishing the relation between $b$ and $\alpha$ for the mapped disk image. 

We begin by defining a right-handed Cartesian coordinate system $(x, y, z)$ with its origin at $\mathrm{O}$, the center of the dense core. The $\mathrm{z}$ axis coincides with the disk's symmetry axis ($\theta=0$), and the $\mathrm{y}$ axis is chosen as the projection of the vector from $\mathrm{O}$ to $\mathrm{O}''$ onto the equatorial plane. The $\mathrm{x}$ axis is then determined by the right-handed rule, with the positive $\mathrm{x}$ axis intersecting a reference sphere (centered at $\mathrm{O}$) at point $\mathrm{X}$. 

Next, we rotate this frame about the $\mathrm{x}$ axis by an angle of $(\pi/2-\theta_0)$, yielding new coordinates $(x, y', z')$
in which the $\mathrm{x}$ axis remains unchanged. This rotation aligns the $\mathrm{y}'$ axis with the line of sight $\mathrm{O}\mathrm{O}''$, so that the $\mathrm{x}\mathrm{z}'$ plane becomes perpendicular to that line of sight (and thus parallel to the screen). Finally, we perform a subsequent rotation about the $\mathrm{y}'$ axis by an angle of $(\pi/2-\alpha)$ to obtain the coordinate system $(x', y', z'')$. 
In this final system, the $\mathrm{x}'\mathrm{z}''$ plane (which is equivalent to the $\mathrm{x}'\mathrm{z}'$ plane) remains parallel to the observer's screen, with the positive $\mathrm{x}'$ axis intersecting a reference sphere at $\mathrm{X}'$. 

We define the point $\mathrm{F}$ as the location where a photon, originally emitted from $\mathrm{S}$, makes its final crossing of the equatorial plane ($\theta=\pi/2$, or equivalently, $z=0$) along its trajectory $\mathcal{C}$. In Fig.~\ref{fig:configuration}, the radial distance of $\mathrm{F}$ is taken to coincide with the radius of the reference sphere. The residual angular deviation $\gamma$ is identified as $\gamma=\angle (\mathrm{FOR})$. Under the assumption that $r_0\gg M$, $\gamma$ asymptotically approaches
\begin{align}
\gamma\simeq \angle (\mathrm{FOY'}),
\label{eq:gammaasym}
\end{align}
where $Y'$ is the intersection of the rotated positive $\mathrm{y}'$ axis with the reference sphere. Owing to the system's spherical symmetry, the photon remains confined to a single orbital plane, which implies that for even $n$, $\gamma=\angle (\mathrm{SOY'})$, and for odd $n$, $\gamma=\pi-\angle (\mathrm{SOY'})$.

Using the asymptotic equivalence from Eq.~\eqref{eq:gammaasym}, we relate $\gamma$ to the observed screen angle $\alpha$ via spherical trigonometry. Noting that $\angle (\mathrm{X}\mathrm{O}\mathrm{X}')=\pi/2-\alpha$, we apply the spherical law of cosines on the spherical triangles $\mathrm{X}\mathrm{X}'\mathrm{F}$ and $\mathrm{X}\mathrm{F}\mathrm{Y}'$ to obtain 
\begin{align}
\cos \gamma&=\cos \varphi \sin \theta_0, 
\label{eq:trigonometry1}
\\ 
\cos \alpha&=\frac{\cos\varphi \cos \theta_0}{\sin \gamma},
\label{eq:trigonometry2}
\end{align}
where $\varphi$ is an azimuthal coordinate value of $\mathrm{S}$. Eliminating $\varphi$ yields 
\begin{align}
\cos \gamma = \frac{\cos\alpha}{\sqrt{\cos^2\alpha+\cot^2 \theta_0}},
\label{eq:gammaalpha}
\end{align}
which clearly shows how $\alpha$ is related to $\gamma$, thereby incorporating the observer's inclination angle $\theta_0$ into the mapping. 

Combining Eqs.~\eqref{eq:Deltavarphi1}--\eqref{eq:deltavarphigamma} with Eq.~\eqref{eq:gammaalpha} yields a complete relationship between the photon's impact parameter $b$ and the observed screen angle $\alpha$. This relation captures both the dynamical and geometrical properties of the photon trajectory, thereby fully determining the mapping from the accretion disk to the observer's screen---and hence, the disk image as seen by a distant observer.

%%%%%
\section{Disk imaging results and gravitational lensing signatures}
\label{sec:5}
%%%%%

\begin{figure}[t]
\centering
\includegraphics[width=11cm,clip]{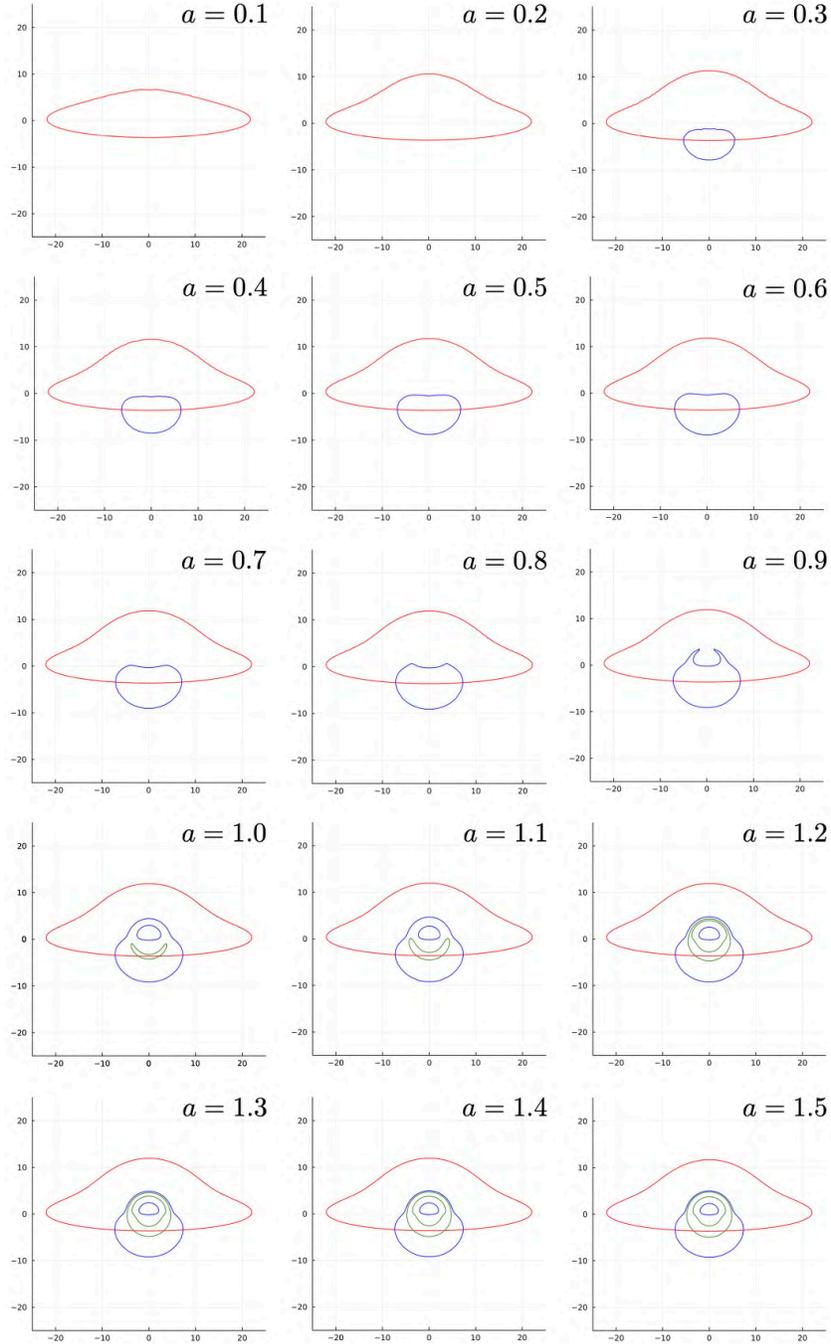}
\caption{Apparent images of the isoradial circle at $r=20M$ in a geometrically thin accretion disk, as seen by an observer at the inclination angle 
$\theta_0=80^\circ$. The isoradial curves are plotted in the $\mathrm{x}''\mathrm{y}''$ plane (the observer's screen coordinates), with units set to $M=1$. The red ($n=0$), blue ($n=1$), and green ($n=2$) curves represent the direct, secondary, and tertiary images, respectively. Higher-order loops $(n\ge 3)$ are omitted for clarity. 
}
\label{fig:theta80}
\end{figure}

\begin{figure}[t]
\centering
\includegraphics[width=12cm,clip]{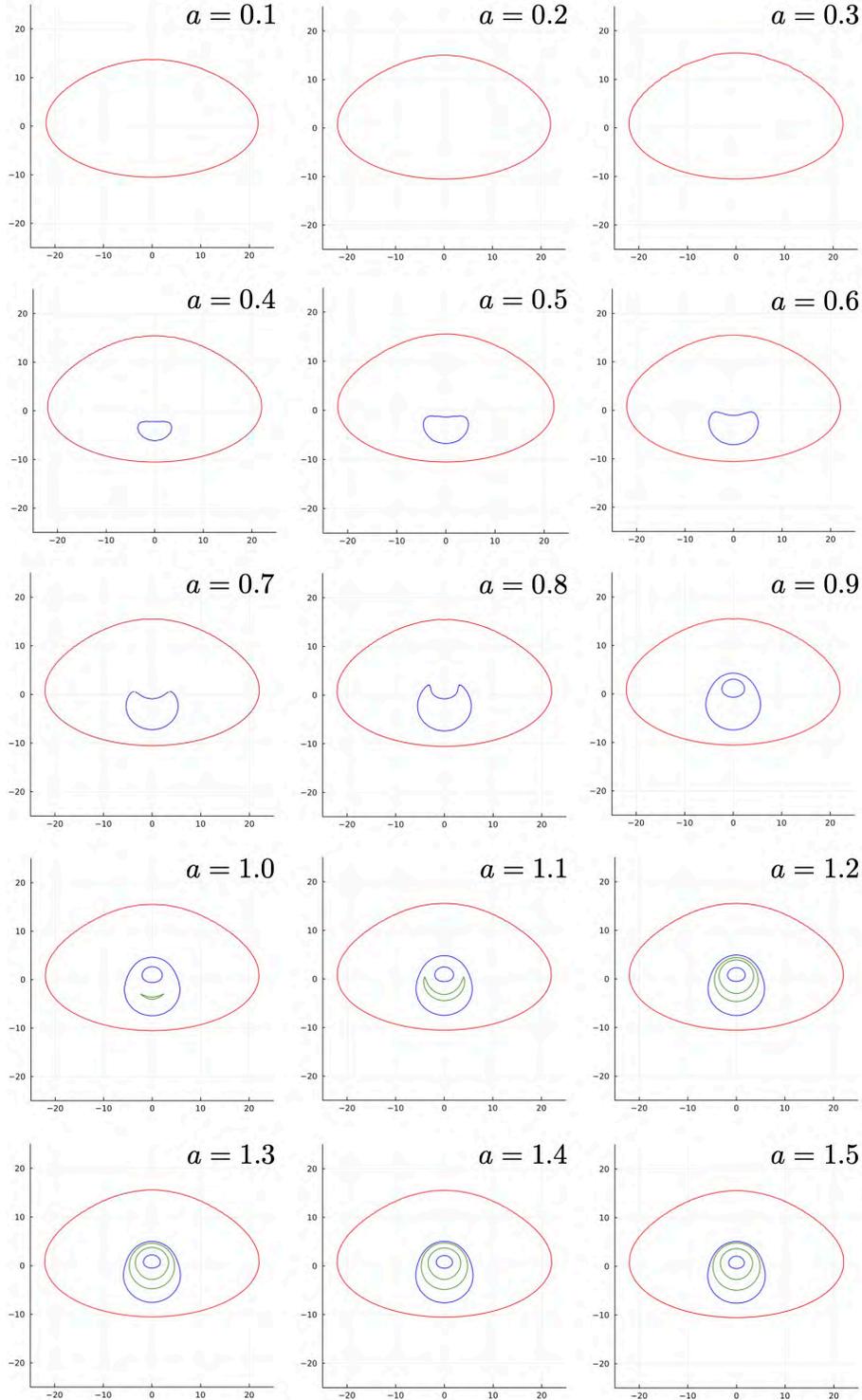}
\caption{Apparent images of the isoradial circle at $r=20M$ in a geometrically thin accretion disk, as seen by an observer at the inclination angle 
$\theta_0=60^\circ$. The isoradial curves are plotted in the $\mathrm{x}''\mathrm{y}''$ plane (the observer's screen coordinates), with units set to $M=1$. The red ($n=0$), blue ($n=1$), and green ($n=2$) curves represent the direct, secondary, and tertiary images, respectively. Higher-order loops $(n\ge 3)$ are omitted for clarity. 
}
\label{fig:theta60}
\end{figure}

Employing the formulations developed in the previous section, we now illustrate the apparent images of an illuminated, geometrically thin accretion disk surrounding the Buchdahl dense core. We focus on a representative source ring at $r=20M$, which exemplifies the characteristic distortions observed in the disk image. We consider inclination angles $\theta_0=80^\circ$ (see Fig.~\ref{fig:theta80}) and $\theta_0=60^\circ$ (see Fig.~\ref{fig:theta60}), with the compactness parameter $a$ ranging from $0.1$ to $1.5$. Note that in this parameter range, the core radius $R$ is smaller than the source ring radius $r=20M$, ensuring that the disk lies well outside the core. For clarity, we classify the images according to the number $n$ of equatorial-plane crossings: $n=0$ corresponds to the primary (direct) image, $n=1$ to the secondary image, and $n=2$ to the tertiary image. We omit isoradial curves with $n\ge3$ in these figures, and their properties are discussed later. In our disk imaging setup, the primary ($n=0$) and secondary ($n=1$) images---characterized by effective deflection angles below $\pi$---appear as direct or mildly lensed images, whereas the tertiary ($n=2$) and higher-order images, which exhibit effective deflection angles exceeding $\pi$, clearly belong to the strong-deflection regime. In each panel, the red isoradial curve ($n=0$) corresponds to the primary image of the source ring, formed by photons that reach the observer without orbiting behind the core. For small values of $a$, this direct image remains nearly elliptical, indicating only mild light bending. As $a$ increases, however, the red curve deviates more markedly from an ellipse, reflecting the enhanced gravitational lensing near the dense core. Moreover, for larger inclination angles (e.g., $\theta_0=80^\circ$), the source ring appears noticeably flatter and more skewed, emphasizing how viewing geometry amplifies the visible distortion.

More prominent features appear in the higher-order images---represented by the blue (secondary image, $n=1$) and green (tertiary image, $n=2$) isoradial curves---which correspond to photons that have crossed the equatorial plane at least once before escaping to the observer. In particular, because the Buchdahl solution allows us to track these curves over a broad range of $a$, we can observe their transitions from very small $a$ to the maximum value considered in our context. Although the following discussion focuses on $\theta_0=80^\circ$ (see Fig.~\ref{fig:theta80}), the corresponding features at $\theta_0=60^\circ$ appear at slightly different values of $a$. When $a \lesssim0.2$, gravitational bending is too weak to produce any visible higher-order loops, such that the observed image is dominated by the direct emission. For values of $a$ between $0.3$ and $0.9$, a secondary image (blue) emerges beneath the main ring, formed by photons bent around the backside of the core. Around $a=0.8$--$0.9$, this secondary loop develops a cusplike feature near its top edge. By $a\approx 1.0$, the blue loop splits into multiple segments, while a crescent-shaped tertiary loop (green) appears in the region enclosed by the secondary loops. Between $a=1.0$ and $1.1$, the tertiary loop gradually grows in size. By $a\approx 1.2$, its cusplike structures merge and then split, ultimately forming two distinct segments. Over the range $a=1.2$--$1.5$, both the secondary and tertiary images exhibit double-loop features. These evolving features highlight the transition of the disk image from a simple ellipse for small values of $a$ to a progressively more intricate structure---with the emergence of higher-order loops---for larger values of $a$, thereby revealing the strong gravitational lensing characteristic of compact Buchdahl cores.

%%%%%
\begin{figure}[t]
\centering
\includegraphics[width=10cm,clip]{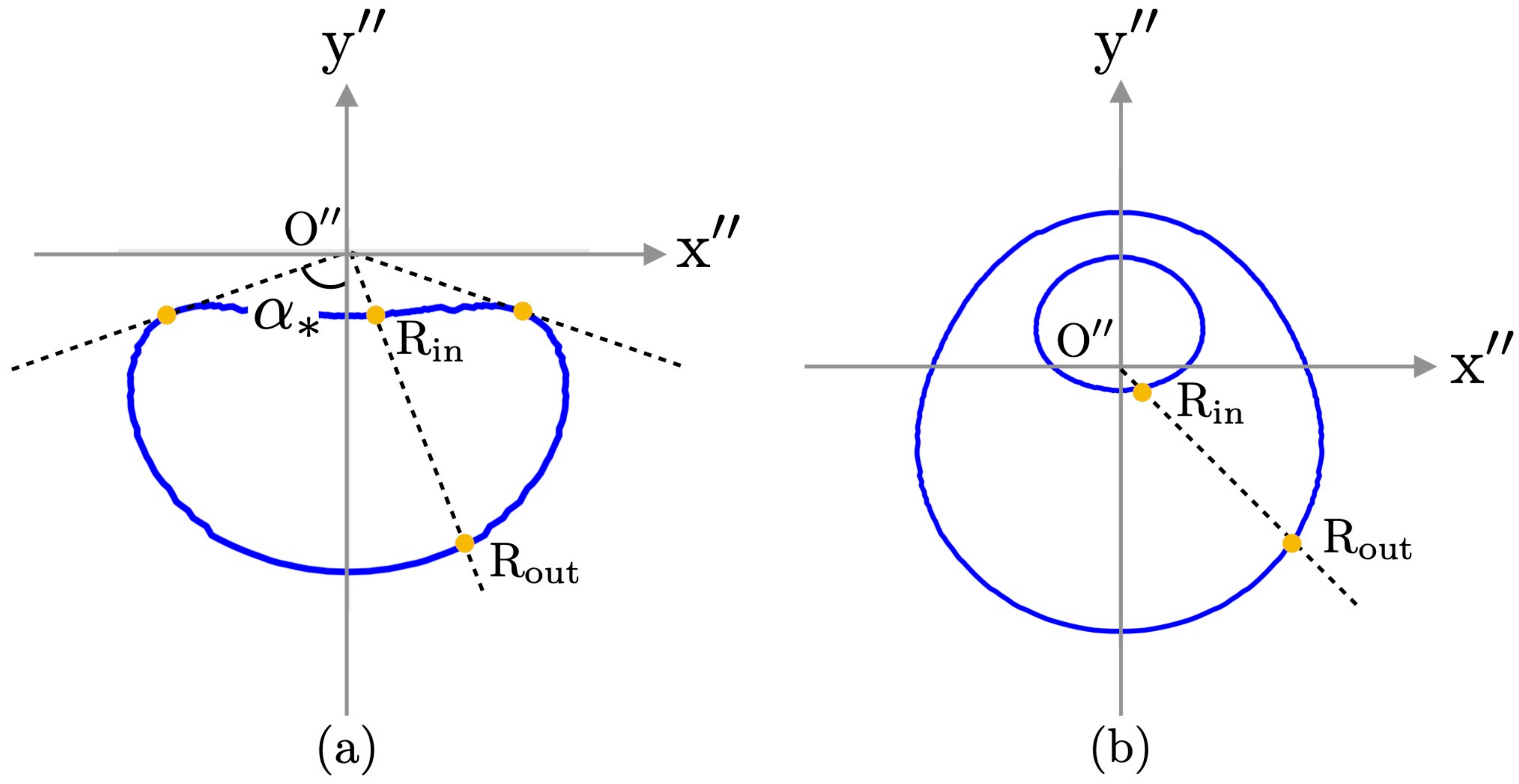}
\\
\caption{Schematic illustration of higher-order images ($n\ge 1$) in gravitational lensing. (a) The image angle $\alpha$ is 
constrained by a cutoff angle $\alpha_*$, causing the isoradial curve to form a closed loop. (b) In contrast, when no such 
constraint exists, the isoradial curve encircles the origin in a double-loop structure. 
In each panel, photons emitted from a specific point on the source ring reach the screen at distinct positions---$\mathrm{R}_{\mathrm{in}}$ for photons that traverse the side closer to the core and $\mathrm{R}_{\mathrm{out}}$ for those arriving from the side farther from the 
core---due to their different impact parameters.
}
\label{fig:alphastar}
\end{figure}
%%%%%
%%%%%
\begin{figure}[t]
\centering
 \begin{minipage}{\textwidth}
    \centering
    \includegraphics[width=10cm,clip]{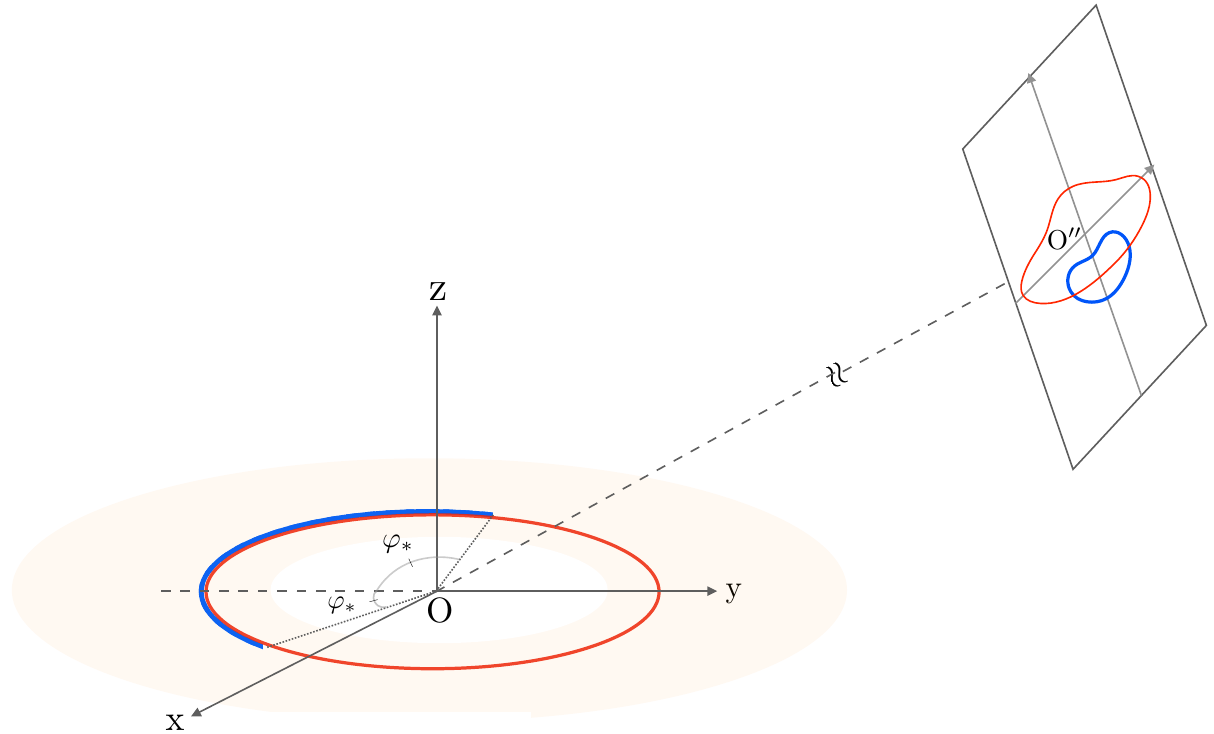}\\
    (a)
  \end{minipage}

  \begin{minipage}{\textwidth}
    \centering
    \includegraphics[width=10cm,clip]{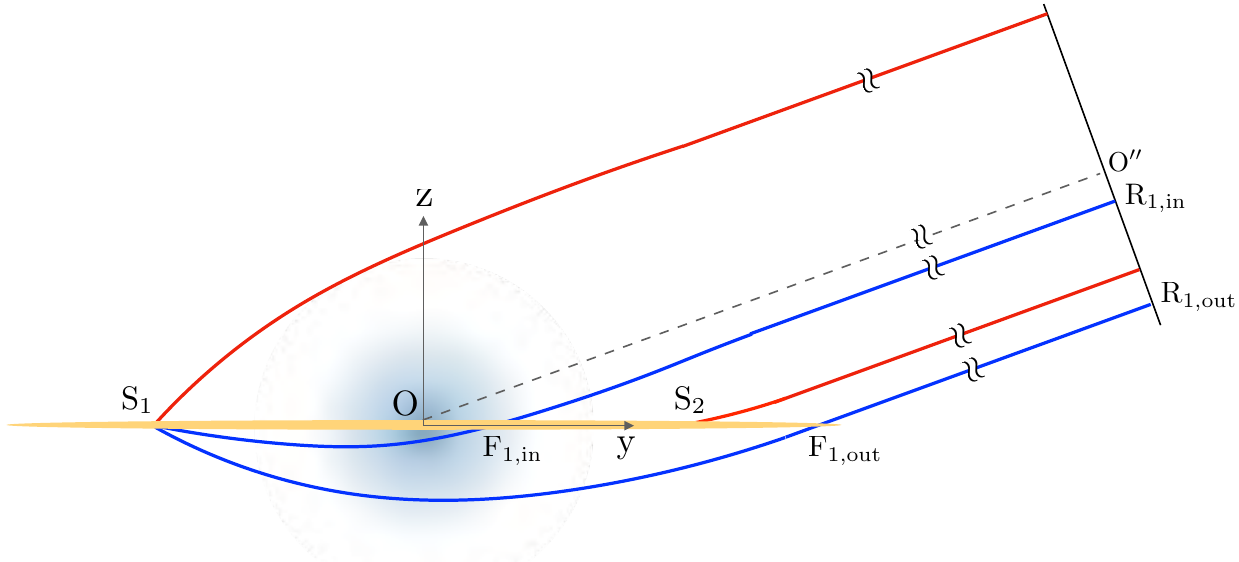}\\
    (b)
  \end{minipage}
\caption{(a) Schematic illustration of the source ranges that produce isoradial curves with a cutoff angle. The red circle (left) indicates the 
entire source ring corresponding to the primary image (red closed curve, right), while the blue arc (left) represents the portion of the source yielding the secondary image (blue closed curve, right). 
(b) Schematic illustration of photon orbits on the $\mathrm{y}\mathrm{z}$ plane: the red curves represent the photon orbits that form the primary image, while the blue curves represent the photon orbits that form the secondary image.}
\label{fig:cutoff}
\end{figure}
%%%%%

One of the key findings from the results is that, as $a$ increases, the earliest signs of higher-order images tend to appear exclusively in the lower half of the screen, with none emerging in the upper half. Consequently, as illustrated in Fig.~\ref{fig:alphastar}(a), we introduce the cutoff angle $\alpha_*$ to denote the boundary in the angular range on the screen such that higher-order images emerge only if
\begin{align}
|\alpha| \le \alpha_*. 
\end{align}
Furthermore, this restriction on $\alpha$ is directly related to the range of the source coordinate $\varphi$. The boundary values of $\varphi$, 
which we refer to as the cutoff source angles, 
are derived from Eqs.~\eqref{eq:trigonometry1} and \eqref{eq:trigonometry2}, 
\begin{align}
\varphi_*=\arccos\left(\frac{\cos \alpha_*}{\sqrt{1-\sin^2\theta_0 \sin^2\alpha_*}}\right).
\label{eq:varphialpha}
\end{align}
When $n$ is odd, only the arc-shaped portion of the source ring satisfying $\pi-\varphi_*\le \varphi \le \pi+\varphi_*$ contributes to the isoradial curves; in contrast, when $n$ is even, only the arc-shaped portion defined by $0\le \varphi\le \varphi_*$ and $2\pi-\varphi_*\le \varphi \le 2\pi$ is relevant (see Fig.~\ref{fig:cutoff}). Table~\ref{table:cutoff} lists the pairs of $\alpha_*$ and the corresponding $\varphi_*$ evaluated for each value of $a$. Thus, only an arc-shaped portion of the source ring contributes to the higher-order isoradial curves observed on the screen. Physically, this cutoff angle $\alpha_*$ arises because the maximum deflection angle of light is bounded when $a < a_0$, preventing photons that require larger bending from reaching the observer. Moreover, for $0\le \theta_0\le \pi/2$, photons that would otherwise appear in the upper half of the screen generally require even larger deflection angles than those appearing in the lower half and therefore often fail to reach the screen. As $a$ increases further within this bounded regime, higher-order images gradually extend into the upper half, even though the deflection angle remains capped by $\alpha_*$.

%%%%%
\begin{table}[t]
  \centering
  \begin{minipage}[t]{\textwidth}
    \centering
    \begin{minipage}[t]{0.28\textwidth}
      \centering
      \begin{tabular}{cccc}
      \hline\hline
      ~~~$a$~~~ & $\mathrm{max}(n)$ & ~~~$\alpha_*$~~~ & ~~$\varphi_*$~~ \\
      \hline
      0.1 & 0 & $180^\circ$ & $180^\circ$ \\
      0.2 & 0 & $180^\circ$ & $180^\circ$ \\
      0.3 & 1 & $68^\circ$ & $23^\circ$ \\
      0.4 & 1 & $79^\circ$ & $41^\circ$ \\
      0.5 & 1 & $84^\circ$ & $59^\circ$ \\
      \hline\hline
      \end{tabular}
    \end{minipage}
    \hspace{0.005\textwidth}
    \begin{minipage}[t]{0.28\textwidth}
      \centering
      \begin{tabular}{cccc}
      \hline\hline
      ~~~$a$~~~ & $\mathrm{max}(n)$ & ~~~$\alpha_*$~~~ & ~~$\varphi_*$~~ \\
      \hline
      0.6 & 1 & $88^\circ$ & $81^\circ$ \\
      0.7 & 1 & $93^\circ$ & $105^\circ$ \\
      0.8 & 1 & $100^\circ$ & $135^\circ$ \\
      0.9 & 1 & $156^\circ$ & $175^\circ$ \\
      1.0 & 2 & $76^\circ$ & $35^\circ$ \\
      \hline\hline
      \end{tabular}
    \end{minipage}
    \hspace{0.005\textwidth}
    \begin{minipage}[t]{0.28\textwidth}
      \centering
      \begin{tabular}{cccc}
      \hline\hline
      ~~~$a$~~~ & $\mathrm{max}(n)$ & ~~~$\alpha_*$~~~ & ~~$\varphi_*$~~ \\
      \hline
      1.1 & 2 & $92^\circ$ & $100^\circ$ \\
      1.2 & 3 & $59^\circ$ & $16^\circ$ \\
      1.3 & 4 & $84^\circ$ & $60^\circ$ \\
      1.4 & $\infty$ & --- & --- \\
      1.5 & $\infty$ & --- & --- \\
      \hline\hline
      \end{tabular}
    \end{minipage}
    \vspace{1ex}
\\
(a) $\theta_0 = 80^\circ$
  \vspace{2ex}
  \end{minipage}
  \begin{minipage}[t]{\textwidth}
    \centering
    \begin{minipage}[t]{0.28\textwidth}
      \centering
      \begin{tabular}{cccc}
      \hline\hline
      ~~~$a$~~~ & $\mathrm{max}(n)$ & ~~~$\alpha_*$~~~ & ~~$\varphi_*$~~ \\
      \hline
      0.1 & 0 & $180^\circ$ & $180^\circ$ \\
      0.2 & 0 & $180^\circ$ & $180^\circ$ \\
      0.3 & 0 & $180^\circ$ & $180^\circ$ \\
      0.4 & 1 & $51^\circ$ & $31^\circ$ \\
      0.5 & 1 & $71^\circ$ & $55^\circ$ \\
      \hline\hline
      \end{tabular}
    \end{minipage}
    \hspace{0.005\textwidth}
    \begin{minipage}[t]{0.28\textwidth}
      \centering
      \begin{tabular}{cccc}
      \hline\hline
      ~~~$a$~~~ & $\mathrm{max}(n)$ & ~~~$\alpha_*$~~~ & ~~$\varphi_*$~~ \\
      \hline
      0.6 & 1 & $85^\circ$ & $80^\circ$ \\
      0.7 & 1 & $99^\circ$ & $107^\circ$ \\
      0.8 & 1 & $125^\circ$ & $144^\circ$ \\
      0.9 & 1 & $180^\circ$ & $180^\circ$ \\
      1.0 & 2 & $38^\circ$ & $22^\circ$ \\
      \hline\hline
      \end{tabular}
    \end{minipage}
    \hspace{0.005\textwidth}
    \begin{minipage}[t]{0.28\textwidth}
      \centering
      \begin{tabular}{cccc}
      \hline\hline
      ~~~$a$~~~ & $\mathrm{max}(n)$ & ~~~$\alpha_*$~~~ & ~~$\varphi_*$~~ \\
      \hline
      1.1 & 2 & $95^\circ$ & $101^\circ$ \\
      1.2 & 3 & $180^\circ$ & $180^\circ$ \\
      1.3 & 4 & $71^\circ$ & $56^\circ$ \\
      1.4 & $\infty$ & --- & --- \\
      1.5 & $\infty$ & --- & --- \\
      \hline\hline
      \end{tabular}
    \end{minipage}
    \vspace{1ex}
\\\
(b) $\theta_0 = 60^\circ$
  \end{minipage}
  \caption{Cutoff angle $\alpha_*$ and corresponding cutoff source angle $\varphi_*$ for isoradial curves 
 exhibiting the maximum number of equatorial crossings, $\mathrm{max}(n)$, as a function of $a$, for a source ring located at $r=20M$. For $a \ge a_0$, $\mathrm{max}(n) = \infty$, and the cutoff angles are undefined (denoted by `---').}
  \label{table:cutoff}
\end{table}
%%%%%

Another key finding is that, even for the same $n$, a single point on the source ring (i.e., having the same $\varphi$) can be mapped to the same screen angle $\alpha$ via two distinct values of the impact parameter $b$. This implies that two distinct photon trajectories---one passing closer to the core center and the other traversing the more distant side---yield the same $\Delta \tilde{\varphi}$. In Fig.~\ref{fig:alphastar}, this is schematically illustrated by the distinct arrival positions: $\mathrm{R}_{\mathrm{in}}$ denotes the arrival point of a photon that has passed on the side closer to the core, while $\mathrm{R}_{\mathrm{out}}$ corresponds to that arriving from the side farther from the core (see also Fig.~\ref{fig:cutoff}). This double-valued mapping highlights the multibranch nature of gravitational lensing near the core and underscores the importance of tracing all possible orbits when analyzing higher-order images. In particular, if 
a cutoff angle $\alpha_*$ exists, each isoradial curve forms a closed loop [see, Fig.~\ref{fig:alphastar}(a)]; by contrast, if $\alpha_*$ does not exist (i.e., the deflection angle is unbounded), the separate isoradial curves encircle the origin, creating a double-loop structure [see, Fig.~\ref{fig:alphastar}(b)]. This distinction shows that the presence or absence of a cutoff angle qualitatively alters the topology of the resulting images, reflecting different regimes of gravitational bending near the core.

\begin{figure}[t]
\centering
\includegraphics[width=12cm,clip]{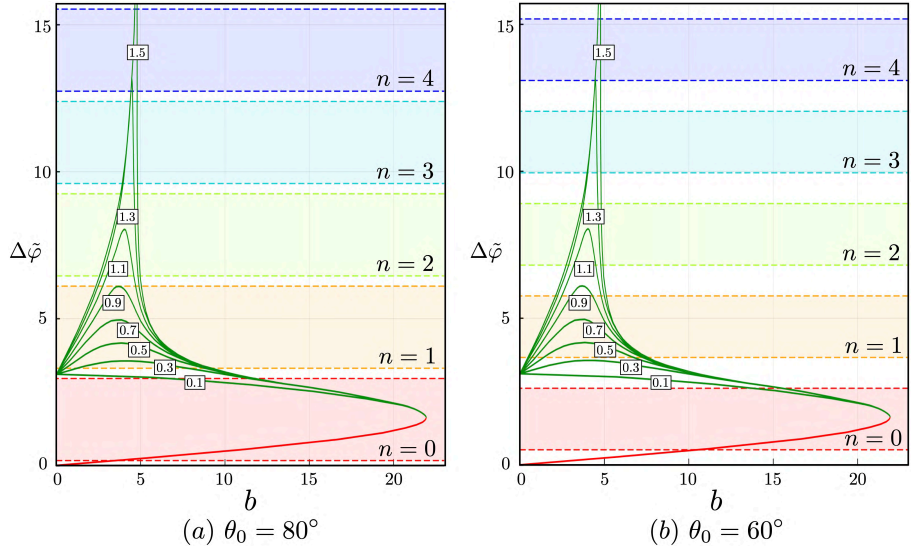}
\caption{
Plots of the total azimuthal change, $\Delta\tilde{\varphi}$, as a function of the impact parameter $b$ for various values of $a$, for a source ring located at $r=20M$, with units set to $M=1$. The colored bands labeled by $n$ represent the range of $\Delta\tilde{\varphi}$ observable on the screen. The red solid curves correspond to photon trajectories without a turning point, while the green solid curves correspond to trajectories that include a turning point.
}
\label{fig:IFD}
\end{figure}

Figure~\ref{fig:IFD} displays plots of the total azimuthal change, $\Delta\tilde{\varphi}$, as a function of the impact parameter $b$ for various values of $a$. In these plots, the color-shaded bands---labeled by $n$---indicate the range of $\Delta\tilde{\varphi}$ observable on the screen. 
Following previous studies~\cite{Gyulchev:2020cvo}, we call these plots the image formation diagram. The red solid curve, obtained from Eq.~\eqref{eq:Deltavarphi1}, corresponds to photon trajectories that do not involve a turning point---i.e., the photons follow a monotonic radial path from the source to the observer. The maximum value of $b$ is determined by the source ring radius (here, $r=20M$). By contrast, each green solid curve, derived from Eq.~\eqref{eq:Deltavarphi2}, corresponds to trajectories that include a turning point.

Focusing on the case $\theta_0=80^\circ$ as shown in Fig.~\ref{fig:IFD}(a) [for $\theta_0=60^\circ$, see Fig.~\ref{fig:IFD}(b)], we examine the detailed behavior of $\Delta \tilde{\varphi}$. For very small $a$ (e.g., $a=0.1$), $\Delta\tilde{\varphi}$ remains confined solely to the $n=0$ band, implying that only the primary image is present. As $a$ takes values between $0.3$ and $0.9$, $\Delta\tilde{\varphi}$ begins to extend into $n=1$ band, signifying the emergence of a secondary image; when the peak of $\Delta\tilde{\varphi}$ lies within this band, a closed single-loop structure on the screen is formed. For $a=1.1$, $\Delta\tilde{\varphi}$ reaches values in the $n=2$ band, giving rise to a tertiary image; at this stage, the $n=1$ band exhibits distinct $b$ ranges for $\Delta\tilde{\varphi}$, indicating a double-loop configuration. As $a$ approaches the critical value $a_0$, the peak of $\Delta\tilde{\varphi}$ increases without bound, resulting in the formation of an arbitrarily large number of higher-order images---even in the absence of a photon sphere. Although strictly speaking, when $a<a_0$ the number of images is finite, for values of $a$ sufficiently close to $a_0$, the effective number of images becomes so large that it is nearly indistinguishable from infinity. For $a > a_0$, however, the presence of a photon sphere causes $\Delta\tilde{\varphi}$ to diverge as $b$ approaches its critical value (as observed for $a=1.5$); this divergence corresponds to an infinite number of higher-order images, all exhibiting a characteristic double-loop structure.

%%%%%
%%%%%

\begin{figure}[t]
\centering
\includegraphics[width=14cm,clip]{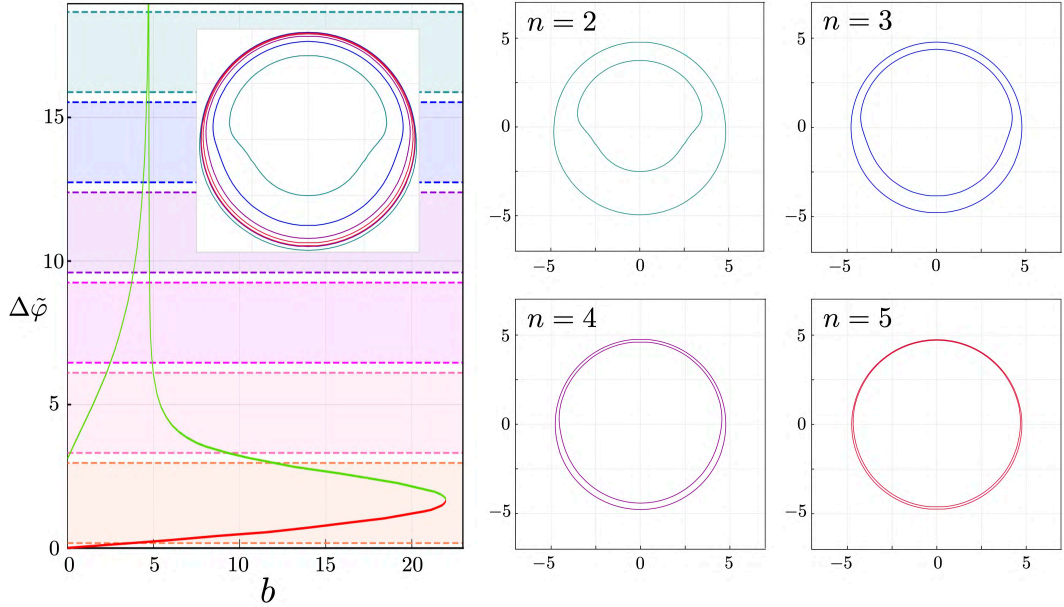}
\caption{
Higher-order isoradial curves for $a=1.5$, $\theta_0=80^\circ$, and $r=20M$ (with $M=1$) are shown in the right panels. In the left-hand panel, the upper figure displays the isoradial curves overlaid in a single plot, while the lower figure presents the image formation diagram.}
\label{fig:HOI_1.5_80}
\end{figure}

\begin{figure}[t]
\centering
\includegraphics[width=14cm,clip]{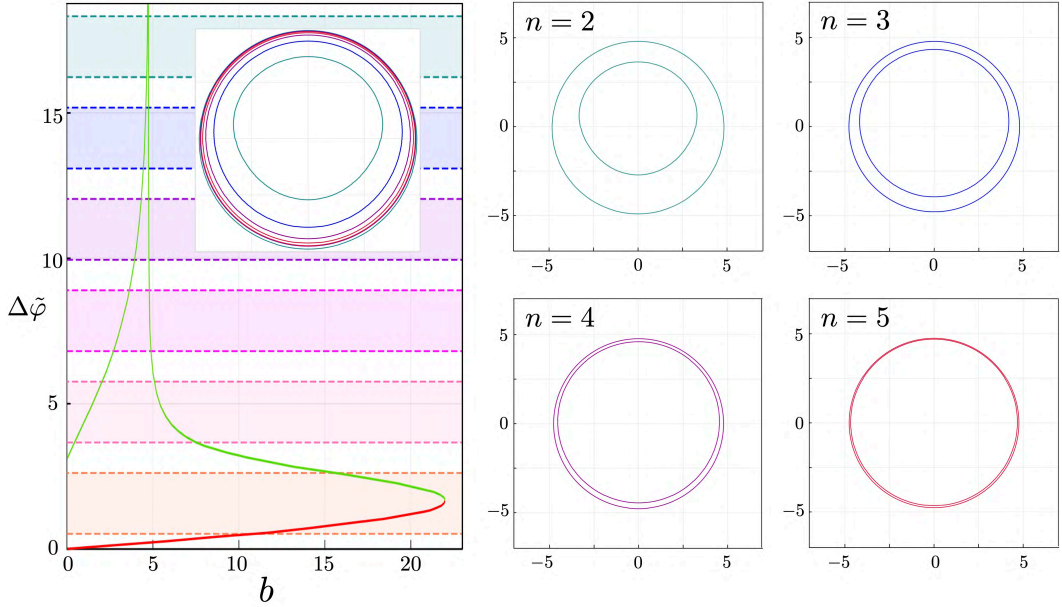}
\caption{
Higher-order isoradial curves for $a=1.5$, $\theta_0=60^\circ$, and $r=20M$ (with $M=1$) are shown in the right panels. In the left-hand panel, the upper figure displays the isoradial curves overlaid in a single plot, while the lower figure presents the image formation diagram.}
\label{fig:HOI_1.5_60}
\end{figure}

Focusing on the case where $a\geq a_0$ (i.e., where an arbitrarily large number of higher-order images emerge), we now examine the detailed behavior of the isoradial curves. The right-hand panel of Fig.~\ref{fig:HOI_1.5_80} displays separate plots of isoradial curves for $n=2$ through $n=5$ at $a=1.5$ and $\theta_0=80^\circ$, while the left-hand panel shows these curves overlaid in a single plot. These panels reveal that, for a given $n$, the double-loop structure associated with the $(n+1)$th image is nestled between the outer and inner loops of the $n$th image. As $n$ increases, the gap between the outer and inner loops within each double-loop structure gradually narrows; by $n=5$, the gap becomes barely discernible, suggesting that as $n$ increases further, the isoradial curves converge in the limit to the single, circular ring corresponding to the photon ring. In contrast to the black hole case, the dense core produces an inner loop generated by photons passing near the core center, and as $n$ increases, this inner loop asymptotically approaches the convergent ring from within. A similar behavior is observed for $a=1.5$ and $\theta_0=60^\circ$, albeit with quantitative differences (see Fig.~\ref{fig:HOI_1.5_60}).

%%%%%
%%%%%

%%%%%
%%%%%

\begin{figure}[t]
\centering
\includegraphics[width=12cm,clip]{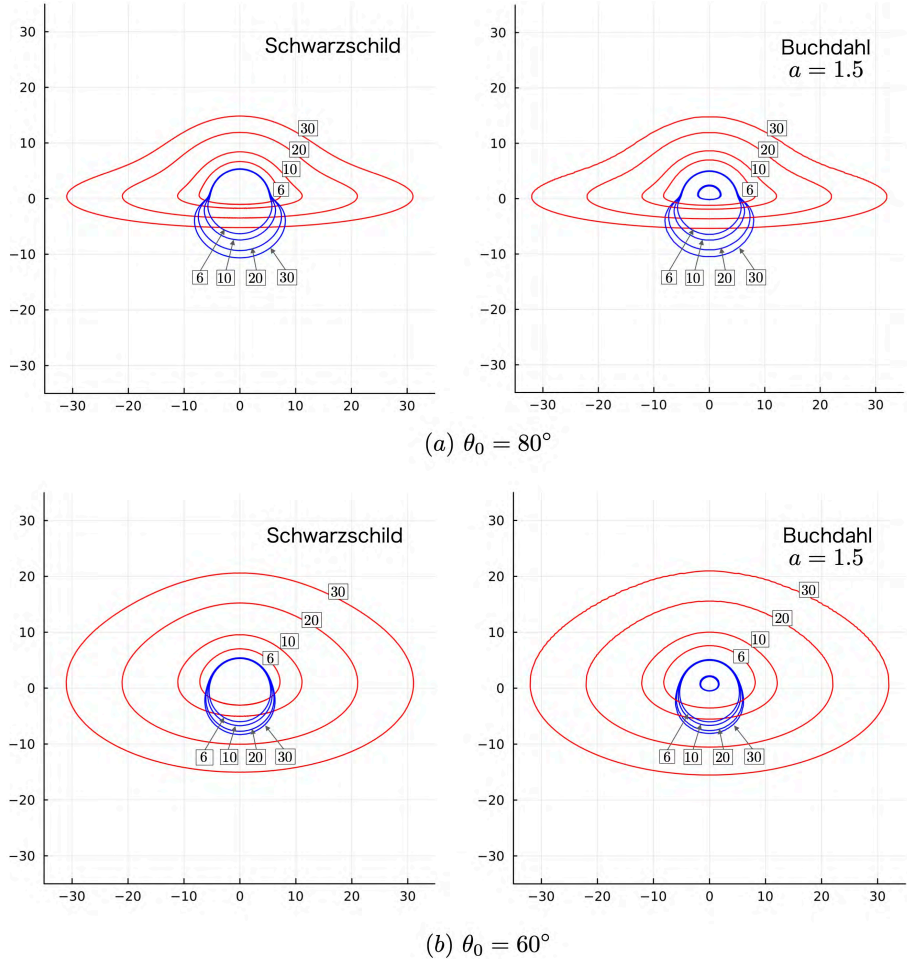}
\caption{
Composite plots of isoradial curves for the Schwarzschild (left) and Buchdahl (right) models with $a=1.5$ under the identical total mass condition (with $M=1$). Curves for source ring radii $r=30M$, $20M$, $10M$, and $6M$ are shown, with the primary images ($n=0$) in red and the secondary images ($n=1$) in blue. Note that in the Buchdahl model the inner loops for $n=1$ are nearly degenerate.}
\label{fig:disk_image}
\end{figure}

Finally, to emphasize the characteristic imaging signatures of the Buchdahl dense core, we present disk images for both the Schwarzschild and Buchdahl models with $a=1.5$ under the identical total mass condition. Figure~\ref{fig:disk_image} displays composite images for source ring radii $r=6M$, $10M$, $20M$, and $30M$ (including images with $n=0$ and $n=1$). The Schwarzschild model exhibits a single-loop structure for each $n$~\cite{Luminet:1979}, whereas the Buchdahl model displays a distinct double-loop structure. Notably, the inner loops of the Buchdahl model remain nearly degenerate across different source ring radii, rendering any geometric variations effectively indistinguishable. This comparison clearly reveals that the double-loop structure is exclusive to the Buchdahl model, thereby underscoring the unique gravitational lensing effects associated with compact objects.

%%%%%
%%%%%

%%%%%
%%%%%

%%%%%
%%%%%

%%%%%%%%%%
\section{Conclusions and discussion}
\label{sec:6}
%%%%%%%%%%
In this paper, we have investigated gravitational lensing and accretion disk imaging in a compact object described by the Buchdahl solution. Starting with a review of the Buchdahl spacetime, we identified the parameter ranges required for a regular and physically viable configuration that satisfies the standard energy conditions. We then derived the equations governing photon orbits, computed the corresponding deflection angles, and constructed a mapping from the illuminated, geometrically thin accretion disk onto the observer's screen---focusing on the isoradial curves corresponding to a representative source ring. Our numerical results reveal the formation of primary and subsequent images with several distinctive features. 

Our study demonstrates that gravitational lensing in the Buchdahl model produces multiple disk images with unique topological characteristics. The primary and secondary images, associated with effective deflection angles below $\pi$, appear as direct or mildly lensed images. In contrast, tertiary and higher-order images emerge only when the effective deflection angles exceed $\pi$, indicating the onset of the strong-deflection regime.

A key finding is the existence of a critical compactness, $a_0$, marking a qualitative transition. For $a < a_0$, the absence of a photon sphere results in only a finite number of images. In this range, the secondary and higher-order images---when present---predominantly display double-loop structures, with each loop capturing the entire source ring. This behavior arises because photons emitted from the same point can reach the screen via two distinct trajectories. Notably, the highest-order image sometimes appears as a single, crescent-shaped loop that does not enclose the screen's center, accompanied by cusplike deformations. This observation implies the existence of a characteristic cutoff angle $\alpha_*$, which eventually restricts the imaged portion of the source ring.
Furthermore, this limited region is captured in duplicate by the two distinct trajectories. Conversely, for $a > a_0$, the existence of the photon sphere allows the deflection angle to grow without bound as the impact parameter approaches its critical value, resulting in an infinite sequence of double-loop structures.

These imaging features---such as double-loop structures and well-defined cutoff angles---appear to be universal signatures of gravitational lensing by compact objects exhibiting central particle scattering. Indeed, similar structures have been reported in disk imaging studies of certain naked singularities featuring central particle scattering (see, e.g., Refs.~\cite{Gyulchev:2020cvo,Gyulchev:2021dvt}), suggesting that these phenomena are not confined to a specific model. In contrast, our analysis based on the Buchdahl spacetime---which describes compact objects without naked singularities---reinforces this universality and demonstrates that such structures can emerge even in the absence of naked singularities. A common property of the underlying spacetimes is the occurrence of central particle scattering, with or without a photon sphere, which suffices to produce a gravitational field strong enough to yield secondary images. An advantage of our model is that the evolution of such structures can be described in a unified manner by a single compactness parameter. Although in the present case the compactness parameter governs the existence of the photon sphere, more generally---regardless of which parameters determine its presence---the corresponding imaging characteristics, such as double-loop structures and cutoff angles, are expected to manifest in a similar manner.

Our results have striking implications for observations. The predicted lensing features---namely, the formation of higher-order images with intricate double-loop structures and the presence of a well-defined cutoff angle $\alpha_*$---provide unique signatures that support dense core models. These features are expected to alter the apparent morphology of accretion disks and photon rings, and they should be detectable with high-resolution imaging. Moreover, because the cutoff angle depends on the source ring radius, the boundaries of the observed disk portion encode valuable geometric information about the underlying spacetime. Preliminary analysis of the brightness profiles and geometric structures in these images suggests subtle deviations from standard black hole predictions. A comprehensive investigation of these aspects is ongoing and will be presented in a forthcoming paper.

\begin{acknowledgments}
This work was supported in part by JSPS KAKENHI Grants No.~JP22K03611, No.~JP23KK0048, No.~JP24H00183 (T. I.), and No.~JP24K00633 (Y. T.) and by Gakushuin University. 
\end{acknowledgments}

%%%
\section*{DATA AVAILABILITY}
%%%
No data were created or analyzed in this study.

\appendix

%%%%%%%%%%
\section{ENERGY CONDITIONS}
\label{sec:A}
%%%%%%%%%%
This appendix analyzes the energy conditions in the parameter ranges $0<|a|<2$ with $k>0$ and $|a|>2$ with $k<0$, where the spacetime is free from curvature singularities. For a detailed discussion of $0<a<2$ with $k>0$, refer to Sec.~\ref{sec:2}. 
%%%
\subsection{$a>2$ and $k<0$}
%%%
In this parameter range, the metric function $f$, as defined in Eq.~\eqref{eq:f}, satisfies $f\in[\:\!a/2,\infty)$.
The energy density $\rho$, given by Eq.~\eqref{eq:rho}, is negative throughout the spacetime ($\rho<0$). 
As a result, the WEC, which requires $\rho\ge 0$, is violated everywhere.

The NEC with $\rho<0$ requires $p/\rho \le -1$.
This inequality with $p/\rho=f/[3(1-f)]$ restricts $f$ to 
$f\in (1, 3/2\:\!]$.
For $2<a\le 3$, the NEC holds for $f\in[\:\!a/2, 3/2\:\!]$
but is violated for 
$f>3/2$. 
For $a>3$, the condition 
$f>3/2$ is always satisfied, leading to a violation of the NEC throughout the spacetime. 

For $\rho<0$, the SEC imposes $p/\rho \le -1/3$, equivalent to $f>1$. 
As $f>1$ always holds for $f\in[\:\!a/2,\infty)$, the SEC is satisfied everywhere. 

The DEC is violated everywhere because $\rho<0$ is incompatible with $\rho \ge |p|$.

%%%
\subsection{$a<-2$ and $k<0$}
%%%
In this parameter range, $f$ satisfies 
$f\in (-\infty, a/2\:\!]$,
and 
$\rho$ remains negative throughout ($\rho<0$). 
This leads to a violation of the WEC everywhere.

The NEC and SEC with $\rho<0$ require 
$f\in (1, 3/2\:\!]$
and $f>1$, respectively.
Since these ranges are incompatible with 
$f\in (-\infty, a/2\:\!]$,
both conditions are violated everywhere. 

The DEC is violated throughout the spacetime because $\rho<0$ is incompatible with $\rho\ge |p|$.

%%%
\subsection{$-2<a<0$ and $k>0$}
%%%
In this parameter range, $f$ lies within the range 
$f\in [\:\!a/2, 0)$
and 
$\rho$ is negative throughout ($\rho<0$). As a result, the WEC is violated everywhere. 
The NEC, SEC, and DEC with $\rho<0$ are all violated due to similar arguments to the case $a<-2$ and $k<0$.

\end{document}